\documentclass[prb,twocolumn,showpacs]{revtex4-1}

\usepackage{graphicx}
\usepackage{multirow}
\usepackage{dcolumn}
\usepackage{bm}
\usepackage[normalem]{ulem}
\usepackage{amsmath}
\usepackage{amsfonts}
\usepackage{amssymb}
\usepackage{color}
\usepackage{colordvi}
\usepackage{dcolumn}

\usepackage[colorlinks=true, breaklinks=true, linkcolor=blue,
urlcolor=blue, citecolor=blue]{hyperref}

\newcommand{\sgn}{\mathrm{sgn}}

\begin{document}

\title{Signatures of quantum phase transitions\\ in the dynamic response of
fluxonium qubit chains}
\date{\today }
\author{Hendrik~Meier$^1$, 
        R.~T.~Brierley$^1$, 
        Angela~Kou$^{2}$, 
        S.~M.~Girvin$^{1,2}$, 
        and Leonid~I.~Glazman$^{1,2}$}
\affiliation{
$^1$Department of Physics, Yale University, New Haven, Connecticut 06520, USA\\
$^2$Department of Applied Physics, Yale University, New Haven, Connecticut 06520, USA
}

\begin{abstract}
We evaluate the microwave admittance of a one-dimensional chain of fluxonium qubits coupled by shared inductors.
Despite its simplicity, this system exhibits a rich phase diagram. 
A critical applied magnetic flux separates a homogeneous ground state from a
phase with a ground state exhibiting inhomogeneous persistent currents. 
Depending on the parameters of the array, the phase transition may be a conventional continuous one, 
or of a commensurate-incommensurate nature. Furthermore, quantum fluctuations affect the transition
and possibly lead to the presence of gapless ``floating phases''. The signatures of the soft modes accompanying the 
transitions appear as a characteristic frequency dependence of the dissipative part of admittance.
\end{abstract}

\maketitle

\section{Introduction}
Vortices of persistent current in superconductors have been viewed for a long time as a testing ground for various models of classical and quantum phase transitions. This is due to relatively strong interactions between vortices coupled with a high degree of control over the vortex arrays. For example, inter-vortex interactions in the presence of a periodic external potential (created experimentally by modulation of the superconducting film thickness \cite{martinoli}) made vortices a convenient target for investigation of commensurability transitions.\cite{pokrovsky,Aubry83} Theoretical studies of the effects of randomness on vortex structure and vortex dynamics have led to the notion of collective pinning,\cite{larkinovchinnikov1,larkinovchinnikov2} with importance stretching well beyond the physics of superconductivity.\cite{fisher_collective} Later, the discovery of high-temperature superconductors triggered studies of the vortex lattice melting transition and glassy behavior\cite{gk,ffh} in layered superconductors as well as structural transitions of vortices interacting with columnar defects.\cite{nelsonvinokur} Because vortices in continuous superconductors have normal cores, vortex motion is dissipative\cite{bardeenstephen} and much of the above-mentioned work used classical statistical mechanics to address collective phenomena in vortex arrays.\cite{blatter_review}

The effects of quantum fluctuations of vortices in continuous films become important only close to the supeconductor-insulator transition, which requires special tuning of the films' normal-state resistance.\cite{goldman_pt} In arrays of Josephson junctions, however, vortices do not have cores, allowing for the study of quantum fluctuations. An array of small superconducting islands connected by Josephson junctions, where the island charging energies were sufficiently large for quantum effects to be important,\cite{fazio} was used in attempts to observe a quantum Kosterlitz-Thouless (KT) transition \cite{haviland} and Mott transition\cite{oudenaarden} in a 1D array. While these works advanced the nanofabrication techniques needed to produce highly regular arrays of small Josephson junctions, the measurement results were ambiguous. The current-voltage ($I$-$V$) characteristics gave inconclusive evidence for a KT transition in single-line 1D arrays.\cite{haviland,haviland2,ootuka} The observations of Ref.~\onlinecite{oudenaarden} were later interpreted to be consistent with a purely classical commensurability transition rather than the quantum Mott transition.\cite{bruder} The suppression of quantum effects in these experiments stemmed from the low inductance of the continuous superconducting wires, which were necessary to make the Josephson junction arrays.

We should note, also, that in the majority of experiments the evidence for the various classical and quantum transitions mentioned above was based on signatures in $I$-$V$ characteristics. This method is limited to addressing highly averaged quantities, and relies on substantial deviations of the investigated system from equilibrium (needed, for example, to overcome the static pinning of vortices).

Recent developments in superconducting qubit techniques offer the possibility of overcoming the described limitations of previous experimental studies of many-body physics of vortices. Typical superconducting qubit experiments address the superconducting system using microwaves.\cite{CircuitQED} The system is only weakly perturbed by the microwave excitation; hence, this spectroscopic approach allows one to probe the system close to equilibrium. On the circuit element side, the development of the fluxonium qubit,\cite{fluxonium} which combines a Josephson junction with a superinductor\cite{gershenson} (i.e., an element exhibiting high inductance and low capacitance) opens avenues for studying quantum effects in superconducting arrays.

\begin{figure}[t]
\centerline{\includegraphics[width=\linewidth]{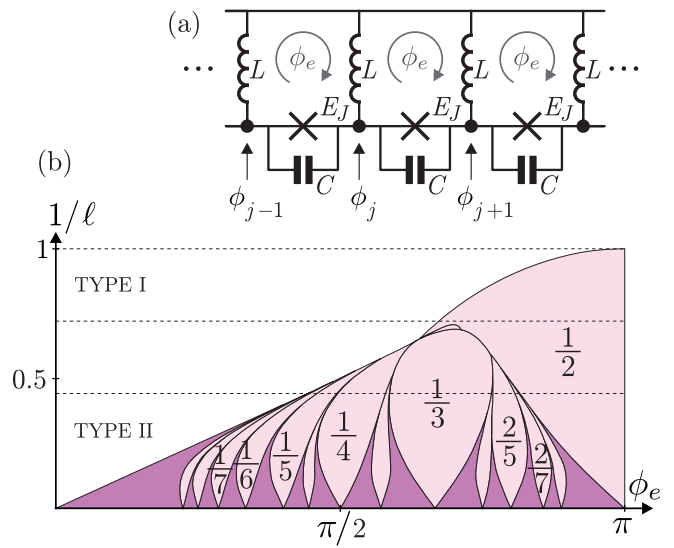}}
\caption{(Color online) 
(a) Quantum circuit of coupled fluxonium qubits threaded by an external flux~$\phi_e$. 
(b) Phase diagram for the classical ground state, cf.\ Refs.~\cite{chiralXY,caille} for
the phase diagram of similar classical models. 
At inductances~$L$ such that $1<\ell\lesssim \sqrt{2}$, the system is in the ``type-I'' regime that
features a single Ising transition from a homogeneous phase into a phase of staggered persistent currents. For
large~$L$ or $\ell\gg 1$ (``type-II'' regime), the system, as a function of external flux~$\phi_e$, develops
subsequent phases of commensurate lattices of vortices or \emph{kinks}.
The rational numbers describe the kink density in the node flux configurations~$\{\phi_j\}$.
Darker shaded regions contain phases with higher denominators that complete the devil's staircase.
The line separating the homogeneous and inhomogeneous phases is given by Eq.~(\ref{phiec_ising}) 
in the type-I and by Eq.~(\ref{x_phiec}) in the type-II regime.
}
\label{fig01}
\end{figure}

In this work, we study theoretically a one-dimensional array consisting of superinductors and a chain of small Josephson junctions as shown in Fig.~\ref{fig01}. 
Two parameters characterize such an array: the ratios of the Josephson ($E_J$) and inductive ($E_L$) energies (see Sec.~\ref{sec:model}), combined 
into a characteristic length
\begin{align}
  \ell = 2\sqrt{E_J/E_L}
\ ,\label{x_ell}
\end{align}
and the ratio~$E_C/E_J$ of charging and Josephson energies, which controls the quantum fluctuations. 
Despite its simplicity, this model allows for a variety of phase transitions of a classical or quantum nature (depending on the ratio $E_C/E_J$)
as a function of the applied magnetic flux~$\phi_e$ per plaquette of the array.

\begin{figure}[t]
\centerline{\includegraphics[width=\linewidth]{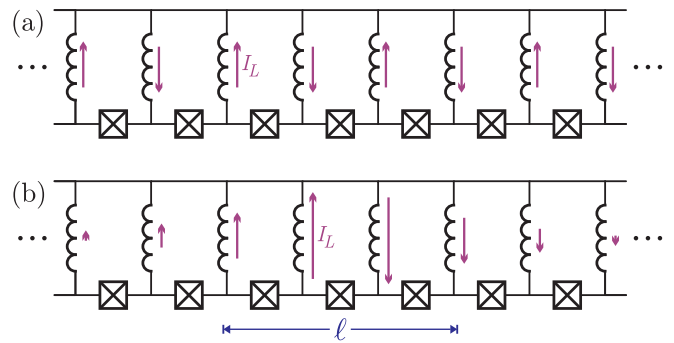}}
\caption{(Color online) 
(a) Staggered persistent currents characterize the high-field ($\phi_e>\phi_e^*$)
ordered phase in the type-I regime.
(b) A vortex of static persistent currents decaying over the length~$\ell$, stable in the
type-II regime.
}
\label{fig_vortices}
\end{figure}

Classically, for fixed $\ell>1$, there is a critical magnetic field $\phi_e^*$ at which the system undergoes a transition 
from a homogeneous state with no persistent currents through the inductors to a state 
with static persistent currents. As a function of the characteristic length~$\ell$, Eq.~(\ref{x_ell}),
we discuss two regimes, cf.\ Fig.~\ref{fig01}(b): a ``type-I'' regime ($1<\ell\lesssim\sqrt{2}$) featuring
a second-order transition to a state of staggered persistent currents as shown in Fig.~\ref{fig_vortices}(a),
and a ``type-II'' regime ($\ell\gg1$), in which the transition takes place by the sequential (first-order) introduction of 
localized vortices of persistent currents in the lattice plaquettes, Fig.~\ref{fig_vortices}(b).
These vortices correspond to \emph{kinks}, meaning discontinuous jumps of height $2\pi$, in the node phase (or node flux)~$\phi_j$, see Fig.~\ref{fig01}(a).
Repulsive interactions between kinks lead to a series of pinned commensurate phases with increasing $\phi_e$.
In the presence of quantum fluctuations, the number of kinks fluctuates and the initial transition at $\phi_e^*$ 
turns into a KT transition. This is followed by commensurate-incommensurate transitions between classical
pinned phases and quantum liquid phases of floating crystalline cells of a kink lattice.

Each of the phases carries a ``fingerprint'' in the microwave absorption spectrum, as the nature of low-energy excitations is sensitive to 
the types of phases and the transitions between them. The microwave spectra also carry information about the crossover from classical to quantum critical behavior in the vicinity of the transition. Yet another advantage of the spectroscopic approach is that it is a linear response to a weak perturbation. In this work, we will highlight the signatures of phase transitions that can be measured using microwave photons.

The paper is organized as follows: In Sec.~\ref{sec:model}, we formulate the mathematical model for the circuit under consideration. 
We also introduce and discuss two methods, capacitative and inductive, for coupling the circuit to an external resonator, 
and give general forms for the radiation absorption rate in the two cases. 

In Sec.~\ref{sec:type-i}, we describe the type-I regime, with $\ell$ comparable to a lattice spacing.
In this limit, the low energy excitations are gapped plasmon oscillations, where the node phases~$\phi_j$ undergo small fluctuations.
The plasmon excitations soften at the critical magnetic flux~$\phi_e^*$, leading to a quantum Ising transition between a phase with the magnetic fluxes $\phi_j=0$ (cf.\ Fig.~\ref{fig01}) everywhere and a staggered phase~$\phi_j = (-1)^j{\bar\phi}$, with order parameter ${\bar\phi}$.

Section~\ref{sec:type-ii} discusses the type-II ($\ell\gg 1$) regime.
Here, the low-energy excitations are associated with the addition or removal of kinks.
As the magnetic field is increased from~$\phi_e=0$, the cost of creating a kink is reduced, leading to a visible peak in the absorption spectrum below the plasmon continuum.
At a critical field, this energy cost vanishes, leading to a proliferation of kinks and the formation of a series of gapped and gapless crystalline phases of kinks.
Each of these has an observable signature in the excitation spectrum for adding or removing kinks. Finally, quantum effects such
as broadening of peaks in the excitation spectrum and the appearance of phases of incommensurate quantum liquids are discussed. 

A discussion of our analysis and its results is presented in Sec.~\ref{sec:discussion}.

\section{Model}
\label{sec:model}

\subsection{Lagrangian}

The circuit in Fig.~\ref{fig01}(a) is described by the Lagrangian
\begin{align}
\mathcal{L}[\phi,\dot{\phi}]
&=
\frac{\hbar^2}{2E_C}\sum_{j=0}^{N-1}\big(\dot{\phi}_{j}-\dot{\phi}_{j-1}\big)^2
- V[\phi]
\ ,\label{x_lagrangian}
\end{align}
where~$E_C=(2e)^2/C$ with $C$ the Josephson junction capacitance, and we formally set $\phi_{j+N}\equiv\phi_{j}$,
assuming periodic boundary conditions. The potential in~(\ref{x_lagrangian}) takes the form
\begin{align}
V[\phi]
&=
\sum_{j=0}^{N-1}\Big\{
\frac{E_L}{2} \phi_j^2
- E_J\big[\cos\big(\theta_j-\phi_e\big)-1\big]
\Big\}
\label{x_potential}
\end{align}
where $E_L=\Phi_0^2/(4\pi^2 L)$, $L$ is the inductance of the inductors and $\phi_e=2\pi\Phi/\Phi_0$ with $\Phi$ denoting the magnetic flux per plaquette and $\Phi_0=h/2e$ being the flux quantum.
Furthermore, we define $\theta_j=\phi_j-\phi_{j-1}$, the phase difference across the Josephson junction between nodes~$j-1$ and~$j$, which
in the following we denote as \emph{link}~$j$.

The potential $V[\phi]$ is not invariant under $\phi_j\to \phi_j+2\pi$, as would be expected for a superconducting system. This is because we neglect the phase slip processes that allow the inductors to relax to the true ground state of the system. Superinductors such as those used in a fluxonium qubit are engineered so that this is a valid approximation.\cite{gershenson}

At large~$C$, the physics is dominated by the (meta)stable configurations~$\{\phi_j\}$ that minimize the static potential~$V[\phi]$. The magnetic flux~$\phi_e$
and the characteristic length scale $\ell$, defined in Eq.~(\ref{x_ell}), are the only parameters in this ``classical'' regime. Throughout
our theoretical analysis, we assume the limit of large systems,~$\ell\ll N$. Note that,
due to the inductive potential, a configuration with local flux~$\phi_j$ is physically distinct from a 
state with flux $\phi_j+2\pi n$ with integer~$n\neq 0$. In the ground state, 
each variable~$\phi_j$ will thus take values between~$-\pi$ and~$\pi$. Furthermore, minimal configurations~$\{\phi_j\}$ necessarily 
satisfy~$\sum_j\phi_j=0$, which physically corresponds to current conservation.

\subsection{Response functions in absorption spectroscopy}

\begin{figure}[t]
\centerline{\includegraphics[width=0.6\linewidth]{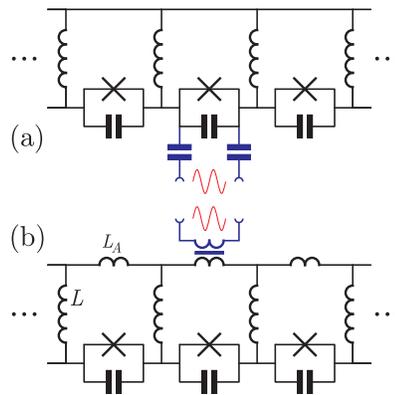}}
\caption{(Color online) 
Schematical antenna setups for absorption spectroscopy: (a) capacitive coupling and (b) inductive coupling
to a microwave resonator.
}
\label{fig_antennas}
\end{figure}

By coupling the circuit in Fig.~\ref{fig01}(a) to microwaves,
signatures of the various phases and phase transition become
observable in absorption spectroscopy. Additional interest
in such experiments may arise because of well-defined peaks
in the absorption spectrum below the plasma gap.
We suggest two, in a certain way complementary, schemes of coupling\cite{CircuitQED}
the circuit to a microwave resonator: (a) inductive coupling and (b)
capacitive coupling, see Fig.~\ref{fig_antennas}. 

\subsubsection{Capacitive coupling}
\label{sec:capacitive}

In the situation of Fig.~\ref{fig_antennas}(a), we add an antenna capacitively coupled to two neighboring nodes~$j_a-1$ and~$j_a$,
which will couple the microwave mode to the Josephson junction at link~$j_a$.

The Hamiltonian for the the microwave mode (with creation and
annihilation operators~$b^\dagger$ and~$b$) and its coupling to
the circuit takes the form
\begin{align}
  \label{eq:antenna_hamiltonian}
  H_{C} = \hbar\omega_R b^\dagger b + g_C(b+b^\dagger)\mathcal{N}_{j_a}
\ .  
\end{align}
Herein, $\omega_R$ is the frequency of the microwave mode and $g_C$ is the coupling constant, which is determined by the impedance of the microwave mode
and the coupling capacitance.\cite{CircuitQED} $\mathcal{N}_{j_a} = i\partial/\partial\theta_{j_a}$
is the Cooper pair number operator for the antenna Josephson junction at link~$j_a$.

Using Fermi's golden rule, the absorption rate for driven radiation with frequencies~$\omega$
incident on the antenna has the general form
\begin{align}
\alpha_{C}(\omega) = \frac{2\pi g_C^2}{\hbar}\sum_m|\langle m|\mathcal{N}_{j_a}|0\rangle|^2\ \delta(\omega-\omega_m)
\label{absorptionrate}
\end{align}
with the sum being over the excited states $|m\rangle$ with energy $\hbar\omega_m$ relative to the ground state.

\subsubsection{Inductive coupling}
\label{sec:inductive}

For the inductive scheme, Fig.~\ref{fig_antennas}(b), we add inductances~$L_A$ into
the former ``ground line'' such that the system remains translationally symmetric.
The antennas themselves may couple inductively to one or several of these inductances.
Here we suppose it solely couples to one link, say~$j_a$.

The inductances~$L_A$ should be chosen small such that the length
\begin{align}
\ell_A=2\sqrt{L/L_A}
\label{ellA}
\end{align}
is not only large itself (in comparison with unity) but also $\ell_A\gg\ell$ (implying $E_{L_A}\gg E_J$), where $\ell$
is the characteristic length defined in Eq.~(\ref{x_ell}).
In this limit, as far as the ground state and excitation energies are concerned, 
the inductances~$L_A$ merely lead to a renormalized characteristic
length~$\ell_{\mathrm{eff}}=\ell/\sqrt{1+(\ell/\ell_A)^2}$. With good
accuracy, we may thus set $\ell_{\mathrm{eff}}\simeq \ell$ and ignore the effects
of the inductances~$L_A$ in the analysis of the model~(\ref{x_lagrangian}).

Using inductively coupled antennas, the microwave modes (described by~$b$ and~$b^\dagger$) effectively
couple to the Josephson phase differences~$\theta_j$ instead of Cooper pair numbers~$\mathcal{N}_j$, cf.\ Eq.~(\ref{eq:antenna_hamiltonian}). 
Currents induced by the antenna at link~$j_a$ decay only over the (large) distance~$\ell_A$, Eq.~(\ref{ellA}). This leads to the following form of the absorption rate:
\begin{align}
\alpha_{L}(\omega) &= \frac{2\pi g_L^2}{\hbar}\sum_m
|\langle m|\bar\theta_{j_a}|0\rangle|^2
\delta(\omega-\omega_m)
\label{absorptionrate_L}
\end{align}
with
\begin{align}
\bar\theta_{j_a} &= 
\frac{1}{N}\sum_k\frac{\hat\theta_k\exp(\mathrm{i}kj_a)}{1+\ell_A^2\sin^2(k/2)}
\simeq\frac{1}{\ell_A}\sum_j \mathrm{e}^{-2|j-j_a|/\ell_A}\theta_{j}
\ ,
\end{align}
where~$k=2\pi n/N$ with $n=0,1,\ldots, N-1$ and $\hat\theta_k=\sum_j\theta_j \exp(-\mathrm{i}kj)$ is the 
Fourier transform of~$\theta_j$. The last approximation is valid in the long-wavelength limit.
The coupling constant $g_L$ in Eq.~(\ref{absorptionrate_L}) is determined by the impedance of the microwave mode
and the coupling inductances.

In contrast to the capacitive coupling, which \emph{locally} couples 
the microwave mode to one Josephson link, the inductive scheme has a much \emph{longer range}~$\ell_A$, Eq.~(\ref{ellA}).
We will specify the spectral absorption rates for the capacitively coupled antenna,~$\alpha_{C}(\omega)$, 
and the inductively coupled antenna,~$\alpha_{L}(\omega)$, for the type-I regime in Sec.~\ref{sec:type-i} and type-II regime in Sec.~\ref{sec:type-ii}.
In the following, we will use units in which~$\hbar=1$.

\section{Type-I regime: Manifestation of a quantum Ising transition}
\label{sec:type-i}

For~$1<\ell\lesssim \sqrt{2}$, as one increases the external flux~$\phi_e$,
the system undergoes a second-order phase transition from
a low-flux homogeneous configuration
$\phi_j\equiv 0$, corresponding to zero persistent current through the inductors, into
a staggered order~$\phi_j={\bar\phi} (-1)^j$ with an alternating
persistent current through the inductors; see Fig.~\ref{fig_vortices}(a). This state
is classically connected to the state with kink density $\rho=1/2$ at larger~$\ell$ in the type-II regime,
cf.\ the phase diagram in Fig.~\ref{fig01}(b).
In this type-I regime, close to~$\phi_e^*$, the relevant low-energy excitations are plasmons, quantized small fluctuations in $\phi_j$.
Mean-field theory, which assumes that these fluctuations are small, predicts
that the transition happens at the critical flux~$\phi_e^*$ given by
\begin{align}
\cos\phi_e^*=-1/\ell^2\quad\textnormal{for}\quad 1<\ell\lesssim \sqrt{2}\ .
\label{phiec_ising}
\end{align}
Technically, it is more convenient to work with the field~$\theta_j=\phi_j-\phi_{j-1}$, which
represents the phase differences over the Josephson links. For~$\phi_e>\phi_e^*$, the field~$\theta_j$
fluctuates around the ordered configuration $\theta_j=2{\bar\phi}(-1)^j$.

\subsection{Mean-field theory}

For~$\phi_e<\phi_e^*$, mean-field theory for the phase differences~$\theta_j$ gives the plasmon spectrum:
\begin{align}
  \varepsilon_k = \frac12\sqrt{E_CE_L}\sqrt{\frac{1}{\sin^2(k/2)}+\ell^2\cos\phi_e}
  \ ,
\label{plasmon-spectrum}
\end{align}
where, in the continuum approximation, the wave numbers are~$k=0,\ldots,2\pi$. The lowest energy states occur for momentum~$k=\pi$. For
momenta~$k$ close to~$\pi$, the spectrum is simplified to
\begin{align}
  \varepsilon_k &\simeq 
  \gamma\sqrt{
   1 + \frac{E_CE_L}{16\gamma^2}(k-\pi)^2
  }  \ .
\label{plasmon-spectrum-soft}
\end{align}
Herein,
\begin{align}
\gamma &= \frac{1}{2}\sqrt{E_CE_L}\sqrt{1+\ell^2\cos\phi_e}
\nonumber\\
&\simeq \sqrt{E_CE_J\sin\phi_e^*}\ |\phi_e-\phi_e^*|^{1/2}
\ ,\quad \phi_e<\phi_e^*\ ,
\label{plasmon-gap}
\end{align}
is the plasmon gap with the second line showing that mean-field theory predicts
the gap closing as a square root as~$\phi_e\rightarrow\phi_e^*$. At the
transition, the softened low-energy plasmon modes have a linear dispersion relation
$\varepsilon_k=u|k-\pi|$, where
\begin{align}
u=\frac{1}{4}\sqrt{E_CE_L}
\label{u}
\end{align}
is the sound velocity.

On the ordered side, for~$\phi_e>\phi_e^*$, 
plasmons describe the fluctuations of the order parameter~${\bar\phi}$.
A similar mean-field theory 
calculation yields for momenta~$k$ with $|k-\pi|\ll 1$ a soft-mode dispersion relation
of the same form as~(\ref{plasmon-spectrum-soft}), with the gap larger by a factor of~$\sqrt{2}$,
\begin{align}
\gamma &= \frac{1}{\sqrt{2}}\sqrt{E_CE_L}\sqrt{1+\ell^2\cos\phi_e}
\ , \quad \phi_e>\phi_e^*\ ,\label{plasmon-gap-ordered}
\end{align}
but otherwise behaves as a function of~$\phi_e-\phi_e^*$ in the same way as the gap~(\ref{plasmon-gap})
on the disordered side.
\bigskip

In absorption spectroscopy, microwave photons excite plasmons so that we expect manifestations
of the critical point at $\phi_e=\phi_e^*$ in the spectral absorption rate~$\alpha_{C}(\omega)$,
Eq.~(\ref{absorptionrate}), or $\alpha_{L}(\omega)$, Eq.~(\ref{absorptionrate_L}), for
capacitive or inductive coupling to the microwave resonator. 
Specifically, we may write 
the Cooper pair number operator~$\mathcal{N}_{j}$ and its conjugate, the phase difference
$\theta_{j}=\phi_{j}-\phi_{j-1}$,
in terms of the plasmon modes,
\begin{align}
  \mathcal{N}_{j} &= \int\frac{\mathrm{d}k}{2\pi}\sqrt{\frac{\varepsilon_k}{2E_C}}(a_k+a^\dagger_k)e^{\mathrm{i}kj}
  \ , \label{eq:plasmon_number} \\
  \theta_{j} &= \mathrm{i}\int\frac{\mathrm{d}k}{2\pi}\sqrt{\frac{E_C}{2\varepsilon_k}}(a_k-a^\dagger_k)e^{\mathrm{i}kj}  
  \ ,    \label{eq:theta}  
\end{align}
where $a_k$ is the annihilation operator for a plasmon at wave number $k$. 
Inserting Eq.~(\ref{eq:plasmon_number}) into Eqs.~(\ref{absorptionrate}), we find 
that at $\phi_e<\phi_e^*$ the spectral absorption rate from the ground state is given by
\begin{align}
  \alpha_C(\omega) &=\frac{\pi g_C^2}{E_C}\ \omega \varrho(\omega)
  \label{fC_plasmons}
\end{align}
for the capacitively coupled antenna, where
\begin{align}
\varrho(\omega) &= \int\frac{\mathrm{d}k}{2\pi}\ \delta(\omega-\varepsilon_k)
\end{align}
is the plasmon density of states. For inductive coupling, inserting Eq.~(\ref{eq:theta}) into Eq.~(\ref{absorptionrate_L}),
we find
\begin{align}
  \alpha_L(\omega) &\simeq \frac{\pi g_L^2 E_C}{\ell_A^2}\ \frac{\varrho(\omega)}{\omega}
    \label{fL_plasmons}
\end{align}
for the softest plasmons with momentum~$k$ near~$\pi$, i.e. $|k-\pi|\ll 1$.
Comparing Eqs.~(\ref{fC_plasmons}) and~(\ref{fL_plasmons}), we infer that the inductive-coupling
scheme leads to stronger response at low energies~$\omega<E_C/\ell_A$ and thus
to higher-contrast results close to the transition at~$\phi_e^*$.

The plasmon spectrum~(\ref{plasmon-spectrum-soft}) implies the density of states
\begin{align}
  \varrho(\omega) &\simeq\frac{1}{\pi u}\frac{\omega\Theta(\omega-\gamma)}{\sqrt{\omega^2-\gamma^2}}
  \ ,
  \label{dos_plasmons}
\end{align}
where $\Theta$ denotes the Heaviside step function. In the frequency region
$\omega-\gamma\ll\gamma$, the absorption rate features a square-root singularity,
\begin{align}
  \varrho(\omega) &\simeq
  \frac{(\gamma/2)^{1/2}}{\pi u}\ \frac{\Theta(\omega-\gamma)}{\sqrt{\omega-\gamma}}
  \ .
  \label{dos_vH}
\end{align}
At the critical field~$\phi_e^*$, where the gap closes, $\gamma=0$, the van-Hove singularity disappears
and Eq.~(\ref{dos_plasmons}) becomes
\begin{align}
  \varrho(\omega) &\simeq \frac{1}{\pi u}
  \ ,\quad \phi_e=\phi_e^*\ ,
  \label{critical_f}
\end{align}
as plasmons have become soft acoustic modes.

\subsection{Fluctuation regime}

Mean-field theory is valid as long as fluctuations are small. Close to 
the critical flux~$\phi_e^*$, plasmon fluctuations become significant as the modes at $k\sim\pi$ soften.
Introducing $\vartheta_j=(-1)^j\theta_j$, we thus write an effective theory
for the ``slow'' field~$\vartheta_j$, keeping only the quadratic leading order in ``discrete gradients''~$\vartheta_j-\vartheta_{j-1}$ so that $16\sum_j\phi_j^2\simeq \sum_j[4\vartheta_j^2+(\vartheta_j-\vartheta_{j-1})^2]$. 
The Euclidean action then reads
\begin{align}
\mathcal{S}
&= \frac{E_J}{2}\int_{-\beta/2}^{\beta/2}\mathrm{d}\tau\sum_j\ \Big\{
\frac{(\partial_\tau\vartheta_j)^2}{E_CE_J}
+\frac{1}{4\ell^2} (\vartheta_j-\vartheta_{j-1})^2
\nonumber\\
&\quad + \frac{1}{\ell^2}\vartheta_j^2 - 2\cos[\vartheta_j-\phi_e(-1)^j]
\Big\}
\label{action}
\end{align}
with~$\beta\rightarrow\infty$ at zero temperature. Outside a Ginzburg region close to the critical field~$\phi_e^*$, one can treat the action~(\ref{action})
in the saddle-point approximation and the mean-field results from the preceding section become accurate.

In order to determine the Ginzburg region, we expand the cosine-potential to fourth order in~$\vartheta_j$
and then employ a continuum approximation, $j\rightarrow x$ and $\vartheta_j-\vartheta_{j-1}\rightarrow\partial_x\vartheta$.
Rescaling coordinates and fields so that they become dimensionless and so the prefactors of the quadratic terms are $1/2$,
we find that close to~$\phi_e^*$ the prefactor of the $\vartheta^4$ term is small if
\begin{align}
|\phi_e-\phi_e^*| \gg \frac{1}{12\ell}\frac{\sqrt{E_C/E_J}}{\sin\phi_e^*}
\ .
\label{ginzburg}
\end{align}
For external fluxes~$\phi_e$ satisfying Eq.~(\ref{ginzburg}), the mean-field results of the preceding
sections are valid.

Very close to~$\phi_e^*$, this Ginzburg criterion breaks down as quantum
fluctuations become strong. The quantum critical behavior, which
is due to the non-linearity of the Josephson current-phase relationship, leads
to an excitation spectrum that is considerably different from mean-field 
theory and should correspond to a $(1+1)$-dimensional quantum Ising chain.\cite{Sachdev}
As a result, e.g., the gap~$\gamma$ in the plasmon spectrum, cf.\ Eq.~(\ref{plasmon-gap}),
is expected to close at~$\phi_e^*$ as
\begin{align}
\gamma \sim \ell^{\frac{1}{2}}(E_J^3E_C)^\frac{1}{4}\sin(\phi_e^*)\ |\phi_e-\phi_e^*|\ ,
\label{gap_linear}
\end{align}
i.e. with critical exponent~$\nu=1$ instead of~$\nu=1/2$ in the
mean-field prediction~(\ref{plasmon-gap}). Furthermore, the $\vartheta^2$-term
in the action~(\ref{action}) will become renormalized, effectively
shifting the critical flux~$\phi_e^*$ to a higher value inside the Ginzburg region.
At criticality, $\phi_e=\phi_e^*$, the system is a liquid with spectrum~$\varepsilon_k=u(k-\pi)$
and plasmon density of states of the form of Eq.~(\ref{critical_f}), but in the fluctuation regime,
the renormalized sound velocity~$u$ has to be considered an effective phenomenological parameter.
While parameters entering the observable quantities~$\alpha_C(\omega)$ and~$\alpha_L(\omega)$, Eqs.~(\ref{fC_plasmons}) and~(\ref{fL_plasmons}),
will be effective ones, the qualitative threshold behavior should still be described in terms of Eqs.~(\ref{dos_vH}) and~(\ref{critical_f}).

Observability of quantum critical behavior requires the system size~$N$ to be larger than
the correlation length at the boundaries of the quantum critical region
as given by Eq.~(\ref{ginzburg}). This leads to the condition
\begin{align}
N\gg \ell^{-1/2}(E_J/E_C)^{1/4}\ .
\end{align}
For typical parameters\cite{typicalvalues} for $E_J$ and $E_C$ (and $\ell\sim 1$ in
the type-I regime), 
the right-hand side of this estimate is of order unity.

\section{Type-II regime: Kinks and quantum phase slips}
\label{sec:type-ii}

Here we consider the limit $\ell\gg1$, corresponding to large inductances $L$. This regime is realizable with superinductors as demonstrated in fluxonium qubits.\cite{fluxonium,gershenson}
The most interesting physical effects are due to the proliferation of \emph{kinks} in 
the node phases~$\phi_j$ corresponding to current vortices as shown in Fig.~\ref{fig_vortices}(b).
These vortices pick up currents over many plaquettes of the order of~$\ell$ and
therefore are stable already at small external fluxes~$\phi_e$.
Thus, phase transitions associated with kink proliferation preempt
the instability driven by fluctuations of plasmons, which remain gapped for all external magnetic fluxes~$\phi_e$.
As a result, the ground state and excitations are fundamentally different from the type-I regime.

We begin with the study of the phase diagram
and the absorption spectrum in the  ``classical'' limit. By ``classical'', we mean
that the capacitances are large enough (and hence $E_C$ is small enough) to make effects due to spontaneous quantum phase slips
negligible but still allow for induced phase slips by microwave absorption.
Quantum effects due to finite capacitances
alter the classical picture and excitation spectrum and will be investigated in Sec.~\ref{sec:quantum}.

\subsection{Classical ground state}
\label{classicalgroundstate}

The classical phase diagram in Fig.~\ref{fig01}(b) is obtained by finding
the configurations~$\{\phi_j\}$ of node fluxes that minimize the potential~$V[\phi]$, Eq.~(\ref{x_potential}),
cf.\ Refs.~\onlinecite{chiralXY,caille}.
They are found from solving the set of equations given by
\begin{align}
0=\frac{1}{E_J}\frac{\partial V}{\partial \phi_j} = \frac{4}{\ell^2}\phi_j
 -\big[&
   \sin(\phi_{j+1}-\phi_j-\phi_e)
  \label{dV}\\   
   &\quad-\sin(\phi_{j}-\phi_{j-1}-\phi_e)
 \big]\ .
\nonumber
\end{align}
Summation over~$j$ of Eq.~(\ref{dV}) yields the constraint~$\sum_j\phi_j=0$, which corresponds to
zero net current to ground.

At zero external flux~$\phi_e$, the ground state of the system is given by the homogeneous
configuration~$\phi_j\equiv 0$, which remains a \emph{local} minimum of $V[\phi]$ for non-zero~$\phi_e$ as long as 
$\cos\phi_e^*>-1/\ell^2$. In the limit of large~$\ell\gg 1$, Eq.~(\ref{dV})
admits non-trivial solutions already at small~$\phi_e\ll 1$ that contain \emph{kinks},\cite{Kardar} static
local configurations of~$\phi_j$ with a jump of the order of~$2\pi$ across one link. In
order to study such soliton solutions, we employ
Villain's approximation\cite{Villain}, in which we expand~$\sin(\theta_j-\phi_e)\simeq \theta_j-\phi_e-2\pi n$
with integer~$n$ such that~$\theta_j-\phi_e-2\pi n$ is small. 

For a \emph{single} kink,
with one jump of $\sim 2\pi$ over the link between sites~$j_0-1$ and~$j_0$ and
$\theta_j=\phi_j-\phi_{j-1}\lesssim \ell^{-1}$ for all $j\neq j_0$, Villain's approach
yields the configuration
\begin{align}
\phi^{\mathrm{kink}}_j &= -\pi\sgn(j+\tfrac{1}{2})\exp\big(-2\big|j+\tfrac{1}{2}\big|/\ell\big)
\ ,\label{x_singlekink}
\end{align}
assuming that~$j_0=0$. 
From Eq.~(\ref{x_singlekink}), we see that~$\ell$, Eq.~(\ref{x_ell}), determines the scale of the width of a kink. Inserting
typical fluxonium values, this width is~$\ell\sim 6$, \cite{typicalvalues}
which sets the smallest system size necessary to observe the physics under discussion.

As the external flux~$\phi_e$ is increased, we may expect the kink solution~$\phi^{\mathrm{kink}}_j$,
Eq.~(\ref{x_singlekink}), to become more favorable than the homogeneous configuration~$\phi_j\equiv 0$
because of the Josephson junction's preference for finite flux gradients at finite~$\phi_e$, cf.\ Eq.~(\ref{x_potential}).
In fact, for the difference~$\Delta=V[\phi^{\mathrm{kink}}_j]-V[0]$ 
between the potential energies of the single kink and homogeneous configuration
we find
\begin{align}
\Delta=2\pi E_J(\phi_e^*-\phi_e)
\ ,\label{Delta}
\end{align}
where
\begin{align}
\phi_e^* &=\frac{\pi}{\ell}
\label{x_phiec}
\end{align}
is the critical external flux in the type-II regime, $\ell \gtrsim 2$, cf.\ Fig.~\ref{fig01}(b).
At flux~$\phi_e^*$, configurations~$\{\phi_j\}$
with kinks become energetically favorable over the homogeneous configuration. 
The critical flux~$\phi_e^*$ is 
analogous to the critical magnetic field~$H_{c1}$ in type-II superconductors,
when vortices begin proliferating.\cite{degennes} Its smallness in~$1/\ell$
reflects that the vortex is able to pick up currents over the (large) length scale~$\ell$,
cf.\ Fig.~\ref{fig_vortices}(b).

For $\phi_e>\phi_e^*$, the density of kinks in the ground state
grows continuously as a function of~$\phi_e$. Each kink, when nucleating individually, brings
an energy gain of $\Delta=2\pi E_J(\phi_e^*-\phi_e)$. On the other hand, this individual
gain has to be balanced with the interaction energy between two kinks.
For the repulsive potential for two kinks at links $i$ and $j$ 
we find
\begin{align}
J_{ij}
&=
\frac{4\pi^2 E_J}{\ell}\exp\Big(-\frac{2|i-j|}{\ell}\Big)
\ ,\label{x_J}
\end{align}
which decays only at large distances on the scale of~$\ell$, Eq.~(\ref{x_ell}).

\begin{figure}[t]
\centerline{\includegraphics[width=\linewidth]{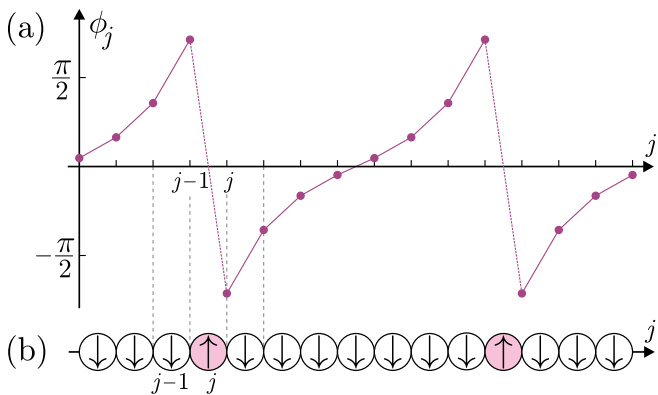}}
\caption{(Color online) (a) Node flux~$\phi_j$ configuration
for a commensurate kink density~$\rho=1/8$ at $\ell=3$. (b) 
The same state represented in the effective spin ladder model for the coupled
Josephson links. Links with upward orientied spin feature
a kink in~$\phi_j$.}
\label{fig04}
\end{figure}

The competition between kink-kink repulsion $J_{ij}$, Eq.~(\ref{x_J}), and
the single-kink energy~$\Delta$, Eq.~(\ref{Delta}), which plays the role
of a chemical potential, completely determines the ground state as a function
of the external flux~$\phi_e$ at $\ell\gg 1$. This interplay may be
effectively described in terms of a classical spin chain model.
Specifically, we associate with each Josephson link~$j$ 
a pseudo-spin that distinguishes
whether there is a kink across it (``spin-up'') or not (``spin-down''),
see Fig.~\ref{fig04}. The effective spin Hamiltonian then has the form
\begin{align}
  H_{\mathrm{cl}} &= \Delta\sum_jn_j
  +\frac12 \sum_{i\neq j}J_{ij}n_in_j
\ ,  \label{x_clHamiltonian}
\end{align}
where $n_j=(\sigma^z_j+1)/2$ and we introduce
Pauli matrices~$\sigma^{x}_j$, $\sigma^{y}_j$, and $\sigma^{z}_j$.

\begin{figure}[t]
\centerline{\includegraphics[width=\linewidth]{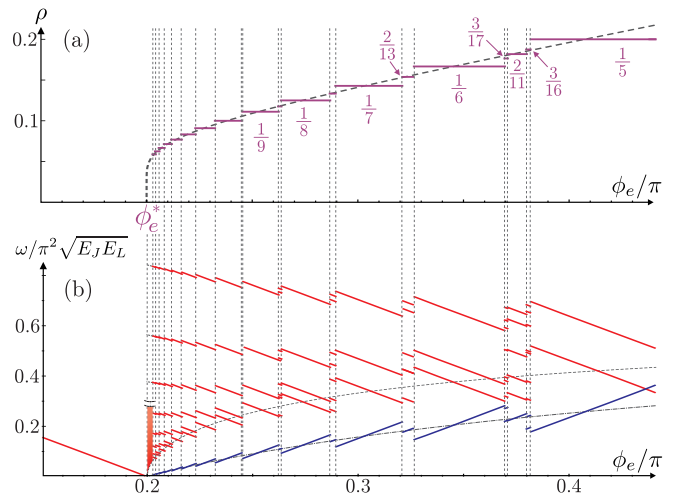}}
\caption{(Color online) 
(a) Kink density~$\rho=q/p$ for the classical ground state as a function of the external flux~$\phi_e$ (for $\ell=5$ fixed). The plot
shows all densities with periodicity~$p\leq 17$; higher-order commensurate phases are
hidden behind the vertical dashed lines. The dashed curve is obtained in the continuum approximation~(\ref{x_kinkdensity}).  
(b) Classical excitation levels by adding (red, falling) or removing (blue, rising) a kink. For fixed~$\phi_e$, the lines
mark the frequencies at which the spectral function exhibits a $\delta$-peak. The upper (lower) dashed
curve shows the minimal energy for adding (removing) a kink in the continuum approximation, cf.\ Eqs.~(\ref{minenergy}) and~(\ref{E_annihilation}).
For excitations close to~$\phi_e^*$, see Fig.~\ref{fig03}.}
\label{fig02}
\end{figure}

The ground states of an Ising chain with infinite-range convex interaction
potentials were systematically studied 
as a function of the ``chemical potential''~$\Delta$ 
by Bak and Bruinsma\cite{BakBruinsma} and Aubry\cite{Aubry83}.
In the language of our model, they showed that in the limit of very large systems $N\rightarrow\infty$,
the kink density~$\rho=\langle n_j\rangle$ depends on~$\phi_e$ in the form
of a devil's staircase:\cite{footnote_1} This means that $\rho$ only takes values in the rational numbers
and rises monotonically and continuously with~$\phi_e$ such that for each rational~$q/p\leq 1/2$, there
is a finite interval in~$\phi_e$ in which $\rho(\phi_e)\equiv q/p$. For our
effective model~(\ref{x_clHamiltonian}), this interval has the width
\begin{align}
\Delta\phi_e (p) \simeq \frac{2\pi p}{\ell^3\sinh^2(p/\ell)}
\label{x_width}
\ .
\end{align}
Figure~\ref{fig02}(a) illustrates the devil's staircase for fixed~$\ell$, showing as a function of~$\phi_e$
the kink density up to commensurability order~$p=17$.

A ground state configuration~$\{\phi_j\}$ 
with kink density~$\rho=q/p$ is periodic,\cite{Hubbard78} $\phi_{j+p}=\phi_j$. For example
the ground state configuration for~$\rho=1/3$ is
$\cdots\uparrow\downarrow\downarrow\uparrow\downarrow\downarrow\uparrow\downarrow\downarrow\cdots$,
with ``primitive cell'' $\uparrow\downarrow\downarrow$. For 
the non-unit fraction $\rho=2/5$, a primitive cell contains a non-trivial basis: $\uparrow\downarrow\uparrow\downarrow\downarrow$.

Figure~\ref{fig02}(a) also illustrates that close to~$\phi_e^*$,
the dependence of the kink density on~$\phi_e$ appears smooth so that we may try to effectively describe
it using a continuum approximation. Physically such a description seems reasonable because close
to~$\phi_e^*$, kinks are sparse such that the large distances between neighboring kinks
``wash out'' the discrete structure of the underlying lattice. Thus, close to
$\phi_e^*$, the ground state density~$\rho$ of the classical model~(\ref{x_clHamiltonian})
is related to the external flux~$\phi_e$ by
\begin{align}
\phi_e(\rho)
&=
\frac{\pi}{\ell}
\Big[
\coth\Big(
\frac{1}{\ell\rho}
\Big)
+
\frac{1}{\ell\rho \sinh^2[1/(\ell\rho)]}
\Big]\ .
\label{x_kinkdensity}
\end{align}
Solving this relation for~$\rho$ leads to the dashed curve in Fig~\ref{fig02}(a).
For $\phi_e-\phi_e^*\gg\phi_e^*$, Eq.~(\ref{x_kinkdensity}) would predict the
relation~$\rho\simeq \phi_e/2\pi$, which in particular implies 
$\rho=1/2$ at half flux quantum, $\phi_e=\pi$, corresponding to the staggered order.
However, discreteness effects are visibly superposed on the continuum model, mostly
due to large intervals in $\phi_e$ that hold the same kink density, cf.\ Eq.~(\ref{x_width}).

Very close to~$\phi_e^*$ such that $\ln[2\phi_e^*/(\phi_e-\phi_e^*)]\gg 1$,
we explicitly find the dependence
\begin{align}
\rho
&\simeq \frac{2}{\ell}\left\{
\ln\Big[
\frac{2\phi_e^*}{\phi_e-\phi_e^*}
\ln\Big(
\frac{2\phi_e^*}{\phi_e-\phi_e^*}
\Big)
\Big]
\right\}^{-1}\ ,
\label{x_kinkdensity_close}
\end{align}
which in particular shows that at $\phi_e\rightarrow\phi_e^*$ the kink density grows with infinite slope.

The transitions between the various commensurate phases of the model~(\ref{x_clHamiltonian}) are first order.\cite{Aubry83}
At zero temperature, quantum fluctuations due to the capacitative term in Eq.~(\ref{x_lagrangian})
facilitate equilibration and prevent hysteresis effects if $\phi_e$ is varied sufficiently slowly in an experiment.
Realistic capacitive energies~$E_C\sim E_J$ lead to extended phases of incommensurate order,
phases of ``floating primitive cells'', around the transitions which we will discuss in Sec.~\ref{sec:quantum}.

\subsection{Classical absorption spectrum}
\label{sec:classicalabsorptionspectrum}

In the regime of kink proliferation, 
the elementary excitations by photon absorption
are the creation of an additional kink or the
annihilation of a kink already in existence
in the ground state. The typical energy scale
associated with such excitations is given by~$E_J/\ell\sim\sqrt{E_JE_L}$.
This scale could possibly already be below the plasmon gap $\sim\sqrt{E_CE_J}$
if $E_L\ll E_C$. At the critical flux~$\phi_e^*$, Eq.~(\ref{x_phiec}),
when the first kinks nucleate in the ground state,
the spectrum for kink excitations extends to zero energy. 
Here, and in the vicinity of~$\phi_e^*$,
these are therefore the relevant low-energy excitations.
On either side of~$\phi_e=\phi_e^*$, the energies correspond 
to isolated low-frequency absorption peaks, see Fig.~\ref{fig03}.

In this section, we discuss the ``classical'' absorption spectrum, i.e.~we neglect spontaneous creation/annihilation of kinks due to quantum phase slips and only
allow for such processes in the context of photon absorption. We discuss how quantum effects alter this picture in Sec.~\ref{sec:quantum}.

\subsubsection{Capacitive vs. inductive coupling}

In a setup with a single \emph{capacitively-coupled} antenna at link~$j_a$ [Fig.~\ref{fig_antennas}(a)],
the system may be excited by \emph{locally} introducing a kink
at the link~$j_a$ of the antenna. In a single absorption act
associated with kink creation or annihilation at link~$j_a$, the system responds 
by boosting the phase difference~$\theta_{j_a}$ over the $j_a$-th Josephson junction
from $0$ to~$2\pi$ or vice versa, whereas phase differences at other links acquire 
negligible changes~$\lesssim 1/\ell$. For this reason, we may estimate the matrix element in the absorption
rate~$\alpha_C(\omega)$, Eq.~(\ref{absorptionrate}), 
in the limit of a single fluxonium qubit\cite{Catelani}, where
the energy difference~$\varepsilon(j_a)$ between the excited and the ground state
is determined by the external flux~$\phi_e$ and the effective potential a kink at~$j_a$ 
feels in the presence of kinks at other sites.

The matrix element entering Eq.~(\ref{absorptionrate}) is then given by
\begin{align}
  \label{eq:x_type2_matrix}
|\langle\theta_{j_a}=2\pi| \mathcal{N}_{j_a}|\theta_{j_a}=0\rangle|
\sim \Gamma/E_C\ .
\end{align}
Herein, the parameter
\begin{align}
  \label{eq:x_phase_slip_amplitude}
\Gamma=\frac{8}{\sqrt{\pi}}(E_J^3E_C)^{1/4}\ \exp\big(-8\sqrt{E_J/E_C}\big)
\ ,
\end{align} 
with dimensions of energy, is the amplitude of a quantum phase slip.\cite{MatveevLarkinGlazman}
In determining matrix elements and absorption rates, we assume the
typical limit $E_J\gg E_C \gg E_L \gg \Gamma$ and $\omega\gg\Gamma$
for microwave frequencies~$\omega$.

As a result, we obtain for the absorption rate in the case of capacitive coupling
\begin{align}
\label{fCkink}
\alpha_C(\omega)
&= 2\pi \tilde{g}_C^2\left(\frac{\Gamma}{E_C}\right)^2\ \delta(\omega-\varepsilon(j_a))
\end{align}
with $\tilde{g}_C\sim g_C$. It features a single peak at frequency~$\varepsilon(j_a)$, which corresponds
to the energy cost for adding or removing a kink from the ground state at link~$j_a$
The spectrum of energies~$\varepsilon(j)$ is determined
in Secs.~\ref{sec:addition} and~\ref{sec:annihilation}, see also Fig.~\ref{fig02}(b).
\bigskip

The \emph{inductive coupling} setup [Fig.~\ref{fig_antennas}(b)], as discussed in Sec.~\ref{sec:inductive},
couples the single antenna to links over the long range of~$\ell_A$. For realistic situations, we may assume 
that $\ell_A$ is much larger than the length~$\ell$, the relevant scale for the statics of kinks. 
This allows the insertion of kinks at any link over a long range.
If the antenna is situated at~$j_a$, the absorption rate~$\alpha_L(\omega)$, Eq.~(\ref{absorptionrate_L}),
becomes
\begin{align}
\alpha_L(\omega) &= \frac{2\pi\tilde{g}_L^2}{\ell_A^2}
  \left(\frac{\Gamma}{\omega}\right)^2
  \sum_j \mathrm{e}^{-2|j-j_a|/\ell_A}\ 
   \delta(\omega-\varepsilon(j))         
\ ,
\end{align}
where~$\tilde{g}_L\sim g_L$, as the matrix element in Eq.~(\ref{absorptionrate_L}) is given by
\begin{align}
  \label{eq:x_type2_matrix}
|\langle\theta_{j}=2\pi| \theta_{j}|\theta_{j}=0\rangle|\sim \Gamma/\varepsilon(j)\ .
\end{align}
For practical situations, because of the long range, it should therefore be sufficient to restrict 
ourselves to kink densities~$\rho=q/p$ with $p\ll \ell_A$.
In this case, the quantum phase slip due to photon absorption may happen at any link
within a ground state primitive cell, and the absorption rate is simply written as
\begin{align}
\alpha_L(\omega) &= \frac{2\pi \tilde{g}_L^2}{\ell_A}
          \left(\frac{\Gamma}{\omega}\right)^2\big[\varrho_+(\omega)+\varrho_-(\omega)\big]
\ ,
\label{fL}
\end{align}
where $\varrho_+(\omega)$ and $\varrho_-(\omega)$ are the spectral functions associated 
with the creation and annihilation, respectively, of a kink. We determine them in Secs.~\ref{sec:addition} 
and~\ref{sec:annihilation}.

\subsubsection{Absorption-induced addition of kinks}
\label{sec:addition}

\begin{figure}[t]
\centerline{\includegraphics[width=\linewidth]{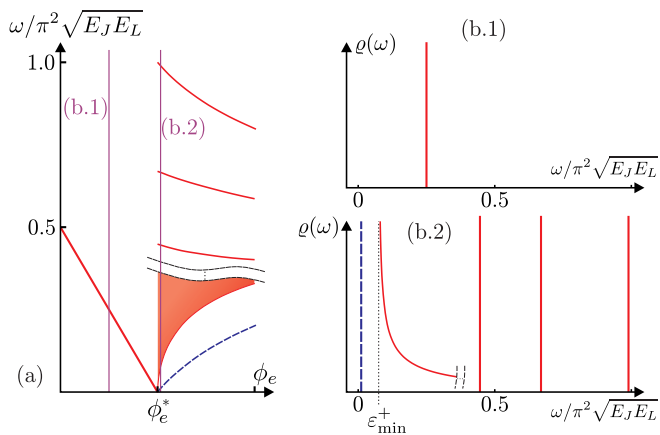}}
\caption{(Color online) 
Classical absorption spectrum close to the critical flux~$\phi_e^*$. 
(a) Photon absorption spectrum as a function of external flux $\phi_e$. Red solid lines correspond to
excitations associated with the addition of a kink, blue dashed lines to removing a kink.
(b.1) At~$\phi_e<\phi_e^*$, cf.\ Fig.~\ref{fig01}(b), there is only one $\delta$-peak in
the spectral function~$\varrho(\omega)$, corresponding to introducing
a kink into the homogeneous ground state. (b.2) For~$\phi_e>\phi_e^*$,
inhomogeneous broadening leads to a quasi-band of excitations related to kink creation,
which for large energies turns into a series of $\delta$-peaks. The lowest excitation is
a single peak (for unit-fraction densities $\rho=1/p$) corresponding to the annihilation
of a kink.}
\label{fig03}
\end{figure}

In this and the next section, we discuss the spectral function
for excitations by adding or removing kinks. We focus mostly
on the situation close to the critical flux~$\phi_e^*$.
Figure~\ref{fig03} shows 
the photon absorption spectrum
as a function of the external magnetic flux~$\phi_e$.

For external magnetic fluxes below the critical value $\phi_e^*$, Eq.~(\ref{x_phiec}),
the ground state is homogeneous, $\phi_j\equiv 0$, and photon absorption
can only result in kink creation. The energy cost of a single kink
is equal to~$\Delta$, Eq.~(\ref{Delta}), so that we find
\begin{align}
\varrho_+(\omega) &= \delta(\omega-2\pi E_J(\phi_e^*-\phi_e))
\quad \text{for}\quad\phi_e<\phi_e^*\ .
\label{spectrumleft}
\end{align}
The absorption frequency decreases linearly as a function of~$\phi_e$ until
it reaches~$0$ at the transition field~$\phi_e^*$.
\bigskip

For~$\phi_e>\phi_e^*$, the ground state of the system 
carries a commensurate kink density~$\rho=q/p$. 
Because of kink-kink repulsion, the energy costs for \emph{adding} 
another kink at a particular site~$j$ is smallest if this link is put 
in the middle between two existing kinks, which are separated by a length
of either the ceiling or floor integer of~$\rho^{-1}$.\cite{Hubbard78}

Let us consider a magnetic flux slightly above~$\phi_e^*$. 
Here, kinks are sparse so that the discreteness of the lattice and 
peculiarities related to higher-order commensurate densities~$\rho=q/p$ with~$q\neq 1$ are less important,
at least for low-energy additional kinks. Since also, according to Eq.~(\ref{x_width}),
most of the $\phi_e$-space is filled by unit fractions~$\rho=1/p$, 
we focus at first on excitations by (low-energy) additional kinks to such ground states. 
Without loss of generality, we may assume that the additional kink
is introduced at link~$j=0$. Then assuming a ground state kink at~$j$ with $j=1,\ldots,p-1$
and for simplicity $p$ even, the additional kink costs the energy
\begin{align}
\varepsilon^+_p(j) &= \Delta + \frac{4\pi^2E_J}{\ell}\frac{\cosh[(2j-p)/\ell]}{\sinh(p/\ell)}
\ .
\label{classicalspectrum}
\end{align}
Herein, the ground state density $\rho=1/p$ as a function of~$\phi_e$ is found from inverting Eq.~(\ref{x_kinkdensity}).
For a given~$\phi_e$, the lowest frequency at which a photon may create a kink is given by
\begin{align}
\varepsilon^+_{\mathrm{min}}
&= \varepsilon_p(p/2)
= \Delta + \frac{4\pi^2E_J}{\ell \sinh(p/\ell)}
\ ,
\label{minenergy}
\end{align}
which corresponds to adding a kink exactly in the center of the primitive cell of the kink lattice.
For $\phi_e$ close above~$\phi_e^*$, using Eq.~(\ref{x_kinkdensity_close}), we find 
\begin{align}
\varepsilon^+_{\mathrm{min}}
&\simeq 
\frac{8\pi^{2}E_J}{\ell}\sqrt{\frac{\phi_e-\phi_e^*}{2\phi_e^*\ln[2\phi_e^*/(\phi_e-\phi_e^*)]}}
\ .
\label{minenergy_close}
\end{align}
Thus, at the transition at $\phi_e^*$,
the gap~(\ref{minenergy}) closes essentially as a square-root law.
The absorption gap~$\varepsilon^+_{\mathrm{min}}$ as a function of~$\phi_e$ is plotted
as a dashed line in Fig.~\ref{fig02}(b) and also as the lower band edge in Fig.~\ref{fig03}(a).

Adding a kink in the proximity of the center of the primitive cell is
energetically least costly. Since (for $p$ even) $\varepsilon^+_p(j)=\varepsilon^+_p(-j)$,
each excitation level is doubly degenerate within one primitive cell. For small~$j-p/2\ll\ell$, we expand Eq.~(\ref{classicalspectrum})
and find
\begin{align}
\varepsilon^+_p(j) &\simeq \varepsilon^+_{\mathrm{min}} + 
\frac{8\pi^2E_J}{\ell^3}\frac{(j-p/2)^2}{\sinh(p/\ell)}
\ .
\end{align}
The accumulation of absorption levels close~$\varepsilon^+_{\mathrm{min}}$ turns
into the usual van-Hove singularity in the continuum approximation, which
is valid for the low lying absorption levels close to~$\phi_e^*$. Explicitly,
for frequencies~$\omega$ slightly above the energy gap, 
$\omega-\varepsilon^+_{\mathrm{min}}\ll\varepsilon^+_{\mathrm{min}}$,	
the spectral function for photon absorption associated with kink creation
reads
\begin{align}
\varrho_+(\omega) &\simeq \frac{\sqrt{\ell\sinh(p/\ell)}}{2\pi(p/\ell)\sqrt{2E_J}}\ \frac{\Theta(\omega-\varepsilon^+_{\mathrm{min}})}{\sqrt{\omega-\varepsilon^+_{\mathrm{min}}}}
\label{rho_continuum}
\end{align}
for $\phi_e>\phi_e^*$. [We recall that by Eqs.~(\ref{x_kinkdensity}) and~(\ref{minenergy}) $p$ is a function of~$\phi_e$.] Close to~$\phi_e^*$,
\begin{align}
\varrho_+(\omega) &\simeq 
\frac{\sqrt{2}}{\varepsilon^+_\mathrm{min}\ln[2\phi_e^*/(\phi_e-\phi_e^*)]}
\frac{\Theta(\omega-\varepsilon^+_{\mathrm{min}})}{\sqrt{(\omega/\varepsilon^+_{\mathrm{min}})-1}}
\label{rho_continuum_close}
\end{align}
with~$\varepsilon^+_\mathrm{min}$ given by Eq.~(\ref{minenergy_close}).

Equation~(\ref{rho_continuum_close}) is valid for frequencies~$\omega$ between~$\varepsilon^+_\mathrm{min}$
and, roughly, $2\varepsilon^+_\mathrm{min}$. Integrating over this interval shows that 
the fraction of available kink-addition states with energies between $\varepsilon^+_\mathrm{min}$
and $\sim 2\varepsilon^+_\mathrm{min}$ is of order $2/\ln[2\phi_e^*/(\phi_e-\phi_e^*)]$.
Close but not too close to~$\phi_e^*$, this can already represent a significant fraction of all available 
excited states associated with one additional kink.
We also note that in the same limit close to~$\phi_e^*$, the level spacing~$|\varepsilon_p^+(j)-\varepsilon_p^+(j-1)|$ remains smaller than the
characteristic energy~$\varepsilon^+_\mathrm{min}$ for the lowest $(\ell/2)\ln\ell$
levels. In terms of energy, this corresponds to all states with energies smaller than~$(\ell/2)\varepsilon^+_\mathrm{min}$.
Both observations indicate that, as $\phi_e\rightarrow\phi_e^*$ the low-energy excitation
spectrum associated with the addition of a kink is suitably described by the continuum
model~(\ref{rho_continuum}) and~(\ref{rho_continuum_close}) over a wide range as compared
to the characteristic scale~$\varepsilon^+_\mathrm{min}$.

At larger energies, however, the continuum approximation underlying Eq.~(\ref{rho_continuum}) eventually breaks down,
even close to~$\phi_e^*$. Specifically, adding a kink
right next to an existing kink determines a highest energy level
\begin{align}
\varepsilon^+_{\mathrm{max}} &= \varepsilon^+_p(1) \simeq \frac{4\pi^2 E_J}{\ell}
\ .
\end{align}
The energy difference~$\varepsilon^+_p(1)-\varepsilon^+_p(2)$ between this highest absorption level
and the second highest, in which the additional kink is placed on a next-nearest link, is of order~$E_J/\ell^2$ and to leading order
independent of~$\phi_e-\phi_e^*$, cf.\ Fig.~\ref{fig02}(b). This
invalidates any approach neglecting the discrete structure of the underlying lattice for finite $\ell$. The absorption spectrum at higher energies can then only be described in terms of a sequence of individual $\delta$-function peaks,
\begin{align}
\varrho_+(\omega) = \frac{1}{p} \sum_{j}\delta(\omega-\varepsilon^+_p(j))
\ ,\label{rho_discrete}
\end{align}
cf.\ Fig.~\ref{fig03}(b.2). Note that the spectral weight of an individual peak of Eq.~(\ref{rho_discrete}) 
decreases linearly with~$1/p$, i.e. upon approaching
the critical flux~$\phi_e^*$ from above, as $2/(\ell \ln[2\phi_e^*/(\phi_e-\phi_e^*)])$.
In particular, the peaks in Eq.~(\ref{rho_discrete}) are much less ``bright''
than the single peak on the homogeneous side, $\phi_e<\phi_e^*$, in Eq.~(\ref{spectrumleft}).
\bigskip

Let us finally discuss the situation of external fluxes away from the transition point, $\phi_e-\phi_e^*\gtrsim \phi_e^*$.
For such external fluxes,
the phase diagram for $\phi_e$ is dominated by phases with densities~$\rho=q/p$ where $p$ is small.
According to Eq.~(\ref{x_width}), the phase with a ground state kink density of, e.g., $\rho=1/3$ extends over an interval
of length $\Delta\phi_e (3) \simeq 2\pi/3\ell$, which is almost as large as the interval of the homogeneous phase at $\phi_e<\phi_e^*$.
For finite~$\ell$, this clearly invalidates the continuum approach entirely, and one can only work with the discrete formula
for the absorption spectrum~(\ref{rho_discrete}). As illustrated in Fig.~\ref{fig02}(b), the absorption
spectrum for the larger unit-fraction kink densities, which dominate the phase diagram on the $\phi_e$ axis,
contains fewer discrete $\delta$-peaks, which on the other hand are individually brighter than the peaks close to~$\phi_e^*$.

Within an interval in~$\phi_e$ of constant kink density, the excitation energies fall linearly as a function of~$\phi_e$, as
a higher magnetic flux favors additional kinks, cf.\ Fig.~\ref{fig02}(b). As a result, transitions between the 
commensurate phases are detectable by jumps of the excitation frequencies seen in absorption spectroscopy.

\subsubsection{Absorption-induced annihilation of kinks}
\label{sec:annihilation}

For external magnetic fluxes above the critical flux, $\phi_e>\phi_e^*$,
the ground state carries a finite density~$\rho$ of kinks whose
annihilation constitutes another excitation in the context of
photon absorption. In fact, excited states due to annihilation
of a single kink are classically the energetically lowest excitations
of the system.

We again first consider a sparse unit-fraction density~$\rho=1/p$
at external flux slightly above $\phi_e^*$. In this situation, the energy cost
to annihilate a kink is
\begin{align}
\varepsilon^-_p &= -\Delta - \frac{4\pi^2E_J}{\ell}\ \frac{2}{\exp(2p/\ell)-1}
\ .
\label{E_annih}
\end{align}
Close to~$\phi_e^*$, using Eq.~(\ref{x_kinkdensity_close}),
we may write the dependence on~$\phi_e$ explicitly,
\begin{align}
\varepsilon^- &\simeq 2\pi E_J (\phi_e-\phi_e^*) 
\Big(1 - \frac{2}{\ln[2\phi_e^*/(\phi_e-\phi_e^*)]}
\Big)
\ ,
\label{E_annihilation}
\end{align}
which is valid as long as $\ln[2\phi_e^*/(\phi_e-\phi_e^*)]\gg 1$.
Note that in contrast to the minimal energy to add a kink, Eq.~(\ref{rho_continuum_close}),
the energy for removing a kink depends on~$\phi_e-\phi_e^*$ in the leading order as a linear function.

The spectral function then reduces to a single $\delta$-peak,
\begin{align}
\varrho_-(\omega) = \frac{1}{p} \delta(\omega-\varepsilon^-_p)
\quad \text{for}\quad\phi_e>\phi_e^*
\ .\label{rho_annihilation}
\end{align}
The spectral weight becomes small close to~$\phi_e^*$, where~$p$ is large. This is in complete
analogy with the fate of the individual $\delta$-peaks in the spectral function~$\varrho_+(\omega)$
for absorption-induced kink creation, cf.\ the discussion of Eq.~(\ref{rho_discrete}). Observing
this peak with a capacitively coupled antenna that affects the system locally
thus seems much more difficult than with an inductively coupled antenna, which is able
to excite remote links.

For $\phi_e$ much larger than $\phi_e^*$, the ground state density becomes larger and 
effects due to the discreteness of the lattice become important also for
the spectral function~$\varrho_-(\omega)$. In particular,
kink densities that are not unit fractions and feature more complicated 
primitive cells of the commensurate order may lead to a kink-annihilation spectrum
with more than one peak. For instance, for the rational density
$\rho=3/7$, a primitive cell is given by~$\uparrow\downarrow\uparrow\downarrow\uparrow\downarrow\downarrow$.
Here, removing the second kink is energetically less costly than removing the first or third, 
resulting in two $\delta$-peaks in~$\varrho_-(\omega)$ with the lower one 
being half as ``bright'' as the upper one, cf.\ Fig.~\ref{fig02}(b) for different examples
of this phenomenon ($\rho=3/17$ and $\rho=3/16$).
\bigskip

In conclusion of this section, the classical absorption spectrum shows 
clear signatures of the phase transitions between the various
phases of commensurate kink configurations and, in particular, the transition
at~$\phi_e^*$, at which the first kinks enter the system, cf.\ Fig.~\ref{fig03}.
The lowest excitations on the left side but essentially also on
the right side of~$\phi_e^*$ are single~$\delta$-peaks.

\subsection{Quantum effects}
\label{sec:quantum}

\subsubsection{Quantum phase slips}

The capacitative term in the Lagrangian, Eq.~\eqref{x_lagrangian}, introduces quantum fluctuations into the system.
Besides plasmons, which below the gap of order~$\sqrt{E_CE_J}$ are frozen, these quantum fluctuations
notably become manifest in the form of quantum phase slips,\cite{MatveevLarkinGlazman}
which change phase differences across any of the Josephson junctions by $2\pi$.
This corresponds to the spontanous creation or annihilation of kinks.
For~$\ell\gg 1$, the necessary adjustments $\lesssim \ell^{-1}$ of fluxes~$\phi_j$
neighboring the location of the kink only make minor contributions to the action of the total quantum phase slip.\cite{Rastelli}
The phase slip amplitude~$\Gamma$, which has already been introduced in Eq.~(\ref{eq:x_phase_slip_amplitude}),
measures the coupling strength between the configuration with a kink at a given link and the one without.
Including this coupling, the effective quantum Hamiltonian in the kink-dominated type-II regime~$\ell\gg 1$
is
\begin{align}
  H &= H_{\mathrm{cl}}
  +\frac{\Gamma}2\sum_j\sigma^x_j
\ ,  \label{x_Hamiltonian}
\end{align}
where~$H_{\mathrm{cl}}$ is the classical non-local Ising Hamiltonian, Eq.~(\ref{x_clHamiltonian}).
The quantum phase slips thus enter in the form of an effective transverse field. As 
a result, in the fluxonium limit~$\ell\gg 1$, the circuit Fig.~\ref{fig01}(a) 
constitutes a realization of a quantum Ising model with non-local interactions.\cite{footnote_1}
We remind the reader that the energy scales typically obey~$E_J\gg E_L\gg \Gamma$.

\subsubsection{Low-frequency peaks}
\label{sec:broadenedpeaks}

The lowest excitations in the classical absorption spectrum in the vicinity of~$\phi_e^*$ 
are single peaks, cf.\ Fig.~\ref{fig03} and Eq.~(\ref{spectrumleft}). Quantum fluctuations broaden these peaks. Not too close to~$\phi_e^*$,
this effect is perturbative and the peaks should still be observable
in the realistic quantum regime.

For $\phi_e <\phi_e^*$, according to the classical analysis in Sec.~\ref{sec:addition},
the lowest excitation corresponds to introducing a single kink into the homogeneous ground state.
The excited state then has energy~$\Delta$, Eq.~(\ref{Delta}), and is $N$-fold degenerate
since the kink could have been excited on any of the $N$ links in the system. As we include quantum 
effects due to the $\sigma^x$-term in the Hamiltonian~(\ref{x_Hamiltonian}), this kink,
assumed at link~$i$, effectively hops through the system: In second order in~$\Gamma$,
the kink either is first annihilated and then nucleates at a (different) link~$j$ or vice versa.
In the first process, the effective hopping parameter is given by $\Gamma^2/\Delta$, in
the second by $-\Gamma^2/(\Delta+V_{ij})$ because in the intermediate state the system
holds two virtual kinks affected by the mutual repulsion~$V_{ij}$, Eq.~(\ref{x_J}).
As a result, hopping is effective over distances 
\begin{align}
\tilde{\ell} = \ell \ln(1/d)
\ ,\label{elltilde}
\end{align}
where~$d=\Delta/(4\pi^2 E_J/\ell)$. The single-kink states thus reorganize
into plane waves and the single excitation peak in Fig.~\ref{fig03}(b) is broadened into a band. For $|i-j|\gg\tilde{\ell}$, the two hopping processes
interfere destructively.

If $\Delta^2\gg \tilde{\ell}\Gamma^2$, we may use perturbation theory 
to find the effective quantum spectrum. To second order in~$\Gamma$, 
the spectrum of excited states with a single kink is
\begin{align}
\varepsilon_q &=
\Delta+\frac{\Gamma^2}{4\Delta}
\Big\{
\ell \ln(1+d^{-1})
- 1
\label{broadenedpeak}\\
&\qquad
+\pi\ell
\Big(
\frac{\sin[\tilde{\ell}q/2]}{\sinh(\pi\ell q/2)}
+ \{q\leftrightarrow 2\pi-q\}
\Big)
\Big\}
\nonumber
\end{align}
with the wave number $q=0,\ldots,2\pi$. In terms of~$\phi_e$,
Eq.~(\ref{broadenedpeak}) remains valid as long as $|\phi_e-\phi_e^*|\gg \tilde{\ell}^{1/2}\Gamma/E_J$.
As~$\phi_e\rightarrow\phi_e^*$, quantum fluctuations lead to drastically different low-energy physics.
In the classical limit, $\Gamma\rightarrow 0$ the spectrum reduces to a single peak at~$\Delta$, cf.\ Eq.~(\ref{spectrumleft}).

The bandwidth of the
dispersive part in Eq.~(\ref{broadenedpeak}) is of order~$\ell\Gamma^2/\Delta$.
However, because the dispersive part is exponentially suppressed for momenta~$1/\ell<q<2\pi-1/\ell$, the band
is mostly flat with most states accumulated at the lower band edge.
For instance, a simple estimate using Eq.~(\ref{broadenedpeak}) shows that more than half of the states 
have an energy less than
\begin{align}
\Delta\varepsilon = \frac{\pi\ell\Gamma^2}{2\Delta}\ \mathrm{e}^{-\pi^2\ell/4}
\label{bandwidth_left}
\end{align} 
away from $\varepsilon_{q=\pi}$. The first term in braces in Eq.~(\ref{broadenedpeak}) represents a global renormalization~$\propto\Gamma^2$
of the classical single-kink excited level, due to interaction with virtually created
and annihilated kinks.
\bigskip

For $\phi_e >\phi_e^*$, the lowest energy single-kink excitation is the annihilation of an existing
kink in the ground state (assumed unit-fraction for simplicity), cf.\ Fig.~\ref{fig03}(b.2) and Eq.~(\ref{rho_annihilation}).
At non-zero~$\Gamma$, the induced hopping of kinks makes their positions subject to quantum uncertainty. As a result,
the effective potential a single kink feels in the many-body background
becomes fluctuating. The variance of this effective potential then 
determines the width of the broadened excitation peak for annihilation
of a single kink.

Perturbation theory in~$\Gamma$ yields a small
broadening of the peak at~$\varepsilon_-$, Eq.~(\ref{E_annihilation}), 
with width
\begin{align}
\Delta\varepsilon_- \sim \ell^{1/2}\Gamma
\ .\label{bandwidth_right}
\end{align}
Note the linear dependence on the quantum amplitude, in contrast to the broadening of the kink excitation spectrum at~$\phi_e<\phi_e^*$, 
cf.\ Eq.~(\ref{bandwidth_left}). The estimate~(\ref{bandwidth_right}) is valid for small~$\Gamma$ and for~$\phi_e$ sufficiently far
away from the critical value~$\phi_e^*$, specifically
as long as
\begin{align}
\Gamma \ll \frac{E_J}{\ell^{3/2}} \frac{(\phi_e-\phi_e^*)^{1/2}}{\ln^{1/2}[2\phi_e^*/(\phi_e-\phi_e^*)]}
\ ,
\end{align}
close to~$\phi_e^*$, and $\Gamma \ll E_J/\ell^{3/2}$ for $\phi_e\rightarrow\pi$.
If this inequality is not satisfied, especially
close to~$\phi_e^*$, effects of the then strong effective hopping 
are more drastic and the single peak for kink annihilation
fades away into the spectrum of a strongly-correlated quantum liquid,
cf.\ the next section. Away from~$\phi_e^*$, however, the peak
should remain intact, owing to the exponential smallness
of~$\Gamma$, Eq.~(\ref{eq:x_phase_slip_amplitude}).

\subsubsection{Quantum phase diagram}

\begin{figure}[t]
\centerline{\includegraphics[width=0.9\linewidth]{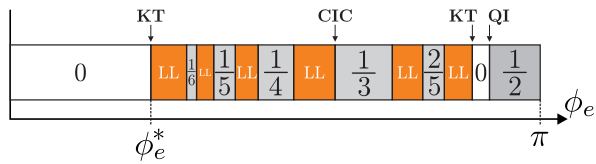}}
\caption{(Color online) 
Sketched quantum phase diagram, cf.\ Ref.~\onlinecite{Garst}, as a function of the external flux~$\phi_e$ for fixed~$\ell= 5$. 
Quantum phase slips, in this illustration with amplitude $\Gamma\sim 0.1 E_J$, melt all commensurate kink densities~$\rho=q/p$ with $p>6$,
leading to extended incommensurate Luttinger liquid (LL) phases, for which
the spectral function has the form of Eq.~(\ref{LLspectrum}). For general~$\Gamma$, Eq.~(\ref{CIC}) determines
the maximum~$p$ such that classical commensurate phases~$\rho=q/p$ survive the quantum fluctuations.}
\label{fig_quantum}
\end{figure}

Quantum Ising models beyond nearest neighbors, i.e. Hamiltonians of the form of Eq.~(\ref{x_Hamiltonian}), 
have recently \cite{Fendley,Garst} been studied theoretically for power-law interactions, 
e.g., in the context of Rydberg atoms \cite{Garst} and experimentally using trapped ions.\cite{IonTraps}
Applying the theoretical results to our model~(\ref{x_Hamiltonian}),
we predict the phase diagram as a function of external flux~$\phi_e$ for fixed~$\ell\gg 1$ and~$\Gamma$ 
to be the one shown in Fig.~\ref{fig_quantum}.

Below a critical flux~$\phi_e^*$, the system is, as in the classical limit, in a homogeneous phase with gapped excitation
spectrum, cf.\ Sec.~\ref{sec:broadenedpeaks}, but then undergoes a Kosterlitz-Thouless (KT) transition
into a gapless phase of ``floating'' kinks. This phase is a Luttinger liquid
and the excitation spectrum for adding or removing kinks is analogous
to the spectrum of adding or removing spinless fermions in a one-dimensional system with strong
repulsion. Upon further increasing~$\phi_e$, the system undergoes a commensurate-incommensurate (CIC) transition
into a ``pinned'' phase featuring a rational kink density~$\rho=q/p$
and a gapped excitation spectrum as in Fig.~\ref{fig02}(b). Quantum effects broaden the ``classical'' peaks according to Eq.~(\ref{broadenedpeak}).
After another CIC transition, the system becomes again a Luttinger liquid. An (even) number of CIC transitions
(depending on~$\Gamma$) may follow before the system undergoes a second KT transition from a liquid
into another gapped homogeneous phase, which in a final quantum Ising (QI) transition turns
into the staggered order corresponding to kink density~$\rho=1/2$.
\bigskip

Let us have a closer look into the phases that feature a finite kink density. For this purpose, 
let us consider a classical ground state with kink density~$\rho=q/p$ and estimate its stability 
to quantum fluctuations. Whereas the adjacent ground state for increasing (decreasing) $\phi_e$ has 
only one extra (fewer) kink, practically all the kinks need to be rearranged in order to minimize the potential energy.\cite{Hubbard78}.
We can interpret this state as the original state of period~$p$ 
with the addition of $p$ defects, in analogy to the two domain walls introduced in an antiferromagnet by flipping one spin.\cite{odintsov,Garst}
For the kink density~$\rho=1/2$, this analogy is perfect.

The quantum $\sigma_x$ term introduces effective hopping of the defects by virtual quasi-simultaneous adding and removing of kinks,
similar to the single-kink hopping discussed in Sec.~\ref{sec:broadenedpeaks}.
The classical energies of the virtually occupied intermediate states are of order $E_J/\ell$, cf.\ Fig.~\ref{fig02}(b), 
giving a hopping amplitude $t\sim \Gamma^2\ell/E_J$.
The kinetic energy gain from such hopping of $p$ defects, which is $\sim p t$, then 
lowers the energy of the states with defects relative to the original period-$p$ ground state
and, as a result, takes away a strip of width~$\sim pt/2\pi E_J$ from both ends of the interval the period-$p$ ground state
has classically occupied in~$\phi_e$. In Fig.~\ref{fig_quantum}, e.g., the $\rho=1/5$ phase 
had to cede almost one half of its classical interval in~$\phi_e$, cf.\ Eq.~(\ref{x_width}), to 
the strips to its left and its right.

In these strips, because of the mobility of the kinks, the commensurate configuration
has melted and been replaced by an incommensurate state of floating
``defective'' primitive cells. This state is gapless and physically described by
a Luttinger liquid. Creation of a kink at a link~$j$ 
is here analogous to injecting a spinless electron at a site~$x$
into a (strongly-coupled) quantum wire.\cite{giamarchi} 
Therefore, using the well-known Luttinger liquid results,
we find for absorption at low energies the spectral function
\begin{align}
\varrho_\pm(\omega) \propto \omega^{\frac{1}{2}(K+K^{-1}-2)}
\ .
\label{LLspectrum}
\end{align}
Within the incommensurate phase, the Luttinger parameter~$K$ varies\cite{Fendley}
as a function of~$\phi_e$. At the CIC transition involving a commensurate kink density~$\rho=q/p$,
it takes the value~$K=1/p^2$, cf.\ Refs.~\onlinecite{giamarchi,Garst}.
If the width of the incommensurate strip exceeds the width of the interval
of the classically commensurate state, i.e. if
\begin{align}
\sinh(p/\ell) \gtrsim \frac{\sqrt{2}\pi E_J}{\ell^2\Gamma} \ ,
\label{CIC}
\end{align}
the classical state with period~$p$ is entirely unstable and merges into an incommensurate
phase of floating defects between two commensurate phases of shorter period~$p$. 
The estimate~(\ref{CIC}) shows explicitly that classical ground states of kink densities
with smaller denominator~$p$ are more immune to quantum fluctuations. In fact, 
the classical phase~$\rho=1/3$, the ``most immune'' pinned phase, would be destroyed entirely only
if $\Gamma$ were so large that it would also destabilize the Luttinger liquid
into a gapped homogeneous quantum paramagnet.\cite{Garst} Based on typical fluxonium parameters \cite{typicalvalues}, we predict that classical phases
of pinned densities~$\rho=q/p$ with~$p\lesssim 10$ survive also in the presence the quantum fluctuations and systems of size larger than~$10$ would be 
needed to observe signatures of Luttinger liquid physics such as an absorption spectrum according to Eq.~(\ref{LLspectrum}) .

Upon decreasing of the external flux to a critical value~$\phi_e^*$, the lowest-$\phi_e$ incommensurate
phase destabilizes into a homogeneous phase in a KT transition\cite{Garst}. 
We note that the value of the critical flux~$\phi_e^*$, because of quantum fluctuations,
will be slightly larger than the classical value from Eq.~(\ref{x_phiec}).
Close to this transition, theory \cite{Garst} predicts a critical~$K=K^*=1/8$.
As the external flux~$\phi_e$ approaches half a flux quantum, 
the system undergoes a similar
KT transition from the rightmost (see Fig.~\ref{fig_quantum}) incommensurate phase (with the same~$K^*$) into
a homogeneous quantum paramagnet. Eventually, there is a quantum Ising transition\cite{Fendley,Garst} to the antiferromagnet-like $p=2$ phase in
the vicinity of $\phi_e=\pi$.

\section{Summary and discussion}
\label{sec:discussion}

\subsection{Summary}

The model circuit, Fig.~\ref{fig01}(a), we have discussed has a surprisingly rich equilibrium phase diagram, cf.\ Figs.~\ref{fig01}(b) and~\ref{fig_quantum}, despite its relatively simple structure.
This arises due to the combination of the nonlinear properties of Josephson junctions and the long-range interactions introduced by the coupling to a common ground.
As a function of the circuit parameters $\ell=2\sqrt{E_J/E_L}$, $E_C/E_J$, and the external field $\phi_e$, the model exhibits 
equilibrium phase transitions of the Kosterlitz-Thouless, commensurate-incommensurate, and Ising classes.
We have shown that circuit QED realizations of this model enable access to low-energy excitations. In particular, 
they allow to identify the quantum phase transitions in linear response by the absorption of microwaves using a capacitively (C) or inductively (L) coupled antenna.
Here we summarize our predictions for the absorption rate~$\alpha_{C/L}(\omega)$ in the 
various characteristic regimes of parameters, particularly $\ell$ and $\phi_e$.

\subsubsection{Type-I regime: $1<\ell\lesssim \sqrt{2}$}

In this limit, the elementary excitations of the system are plasmons. 
The absorption rates~$\alpha_{C}(\omega)$ and~$\alpha_{L}(\omega)$
are given by Eq.~(\ref{fC_plasmons}) and~(\ref{fL_plasmons}), respectively, with the
density of states of plasmon excitations given by Eq.~(\ref{dos_plasmons}). At the critical
external flux~$\phi_e=\phi_e^*$, classically given by Eq.~(\ref{phiec_ising}), the spectrum is gapless 
with a uniform density of states [Eq.~(\ref{critical_f})].
In the proximity of~$\phi_e^*$, where the Ginzburg criterion~(\ref{ginzburg}) is violated
and quantum fluctuations are strong, the gap in the plasmon spectrum grows linearly as a function of
the distance to~$\phi_e^*$, [Eq.~(\ref{gap_linear})], while outside
the Ginzburg region, the classical square root dependence on the external flux [Eqs.~(\ref{plasmon-gap}) and~(\ref{plasmon-gap-ordered})] sets in.

\subsubsection{Type-II regime: $\ell\gg 1$}

This regime, with $E_J\gg E_L$, is closer to the the parameters realized in fluxonium qubits.\cite{fluxonium}
The ground state and excitation spectrum at large flux, $\phi_e>\phi_e^*$ with~$\phi_e^*$ given by Eq.~(\ref{x_phiec}), 
are significantly different from the small-flux regime, $\phi_e<\phi_e^*$, where the ground state is homogeneous.
The elementary excitations are associated with the addition or removal of localized kinks (or vortices).

The absorption spectrum is different for the capacitive [Eq.~(\ref{fCkink})] or inductive [Eq.~(\ref{fL})] coupling of the antenna,
the former adding or removing kinks locally, the latter over an extended range. In the realistic quantum picture,
the lowest excited states in the regime $\phi_e<\phi_e^*$ appear above a gap of order~$\Delta\propto \phi_e^*-\phi_e$
and form a very flat quantum band [Eq.~(\ref{broadenedpeak})]. The band slightly broadens the classical peak in the absorption rate 
[Eq.~(\ref{spectrumleft}) and Fig.~\ref{fig03}(b.1)].
As $\phi_e\rightarrow\phi_e^*$, the gap closes and, upon undergoing a Kosterlitz-Thouless (KT) transition at $\phi_e^*$,
the system enters the phase of a ``floating'' crystal of kinks. The excitation spectrum here at $\phi_e>\phi_e^*$ is of Luttinger-liquid type and
at low energies is given by Eq.~(\ref{LLspectrum}). 

As $\phi_e$ is increased further, depending on the strength of quantum fluctuations
(phase slips), the system undergoes a commensurate-incommensurate (CIC) transition, after which the ground state carries
a pinned classical kink density. Here, the excitation spectrum is essentially the classical spectrum of Figs.~\ref{fig02}(b) and \ref{fig03}(b.2),
where quantum fluctuations slightly broaden the $\delta$-peaks according to Eq.~(\ref{bandwidth_right}).
The classical spectral function for kink addition is given by Eq.~(\ref{rho_continuum_close}), for~$\phi_e$ not too far from~$\phi_e^*$,
and in general by Eq.~(\ref{rho_discrete}). For kink annihilation, the spectral function is displayed in Eq.~(\ref{rho_annihilation}).
For kink densities~$\rho=q/p$ with $q=1$ or $q=2$, which dominate the phase diagram, there is only a single peak in the absorption 
spectrum related to kink annihilation, which is the lowest excitation energy of the system, cf.\ Eq.~(\ref{E_annih}) and 
Figs.~\ref{fig02}(b) and \ref{fig03}(b.2). 

As the external flux is further increased, classical and floating phases alternate as schematically indicated in Fig.~\ref{fig_quantum}.

\subsection{Discussion}

In our theoretical analysis, we built on the availability of a ``superinductance'' such as in fluxonium qubits\cite{fluxonium} 
when envisioning experimental realizations.
As a result, the type-II regime of small $E_L$ could be reached without considering the effects of additional parasitic 
capacitances to ground that would be unavoidable in a realization 
using ordinary electromagnetic inductance.
Such additional parasitic capacitances would suppress quantum fluctuations and thus enhance the classical behavior of the system.
In the type-I regime, it decreases the effective $E_C$ for the low energy plasmon modes, narrowing the quantum critical region for the Ising transition.
In the type-II regime, capacitance to ground decreases the quantum phase slip rate $\Gamma$,\cite{Rastelli} which helps to stabilize the classical pinned phases.

Throughout, we have used periodic boundary conditions for theoretical convenience.
Realistically, it is easier to create an array with open boundary conditions and so for small system sizes there will be edge effects.
These will extend over a length $\sim\ell$, Eq.~(\ref{x_ell}), as this is the scale for interactions in the system. Yet for large systems
whose size significantly exceeds at least the length~$\ell$, our results should be directly applicable.
We have also assumed the temperature is zero.
Both finite temperature and the finite size of the system prevent the formation of long-range order and in principle mean that there will not be a true phase transition.
However, the equilibrium state and spectrum will retain signatures of the infinite system behavior on short enough length and timescales.

We note that there is an intermediate regime $\ell\sim 2$, where the classical model has an incomplete staircase of first order transitions, cf.\ Fig.~\ref{fig01}(b)
and Refs.~\onlinecite{chiralXY,caille}.
In this parameter region, the type-II regime
commensurate and incommensurate phases of finite kink density for~$\phi_e>\phi_e^*$ turn into 
the single Ising phase of the type-I regime. The classical theory\cite{chiralXY,caille} predicts a sequence
of ``superdegenerate''\cite{chiralXY,caille} and multicritical points as the transition at~$\phi_e^*$ changes from first order at large~$\ell$ to second order
in the type-I regime.
The construction of an effective model for this region that also allows one to analytically study the effect of quantum fluctuations
appears difficult. However, the study of the ground states and excitations may be amenable to numerical techniques such as quantum Monte Carlo as the problem is bosonic.

In the type-II regime at large~$\ell$, the low-energy behavior is well described in terms of localized kinks.
Using the locally-coupled capacitative antenna configuration, kinks can be selectively introduced or removed on individual sites.
If multiple kinks are introduced, they will interact over the large length scale $\ell$ according the Hamiltonian~(\ref{x_clHamiltonian}).
This introduces the possibility of investigating 
the many-body non-equilibrium physics of the system.
In the presence of dissipation, it may be possible to create a model system with a driven-dissipative steady-state of interacting kinks, 
an interesting addition to the set of non-equilbrium many body models that can be simulated with circuit QED systems.\cite{houck}

\acknowledgments

We would like to thank G. Rastelli, I. Pop, M. H. Devoret, J. Keeling, Yu Chen, C. Neill, S. Hacohen-Gourgy, and V. Ramasesh 
for interesting and helpful discussions.
This work was supported by NSF DMR--1301798 and DMR--1206612, and ARO W911NF--14--1--0011.
H. M. and R. T. B. acknowledge the Yale Prize Postdoctoral Fellowship.


\begin{thebibliography}{99}
\bibitem{typicalvalues}
For estimates, we use the values~$E_J\approx 10\ \mathrm{GHz}$, $E_L\approx 1\ \mathrm{GHz}$,
and~$E_C/8=e^2/(2C)\approx 3\ \mathrm{GHz}$.
\bibitem{footnote_1}
In order to avoid any confusion, we point out that although our interaction~$J_{ij}$, Eq.~(\ref{x_J}), falls exponentially and so is certainly short-ranged in the sense of scaling, 
it is nevertheless non-zero at an arbitrarily large distance $|i-j|$. This implies
the theoretical completeness of the devil's staircase in our model, even though the ``stairs'' for kink densities $\rho=q/p$ with large~$p$ occupy exponentially small intervals
in~$\phi_e$, cf.\ Eq.~(\ref{x_width}).
\end{thebibliography}
\end{document}